\documentclass[12pt,a4paper]{article}

\usepackage{geometry}
\usepackage{amssymb,amsmath,mathrsfs}
\usepackage{amsfonts}
\usepackage[T1]{fontenc}
\usepackage{graphicx}
\usepackage{color}

\newcounter{mnotecount}%[section]

\newcommand{\mnotex}[1]%{}
{\protect{\stepcounter{mnotecount}}$^{\mbox{\footnotesize $\bullet$\themnotecount}}$ 
\marginpar{%\color{red}%
\raggedright\tiny\em
$\!\!\!\!\!\!\,\bullet$\themnotecount: #1} }

\def\B{\mathscr{B}}
\def\E{\mathscr{E}}

\def\be{\begin{equation}}
\def\ee{\end{equation}}
\def\ben{\begin{equation*}}
\def\een{\end{equation*}}
\def\bea{\begin{eqnarray}}
\def\eea{\end{eqnarray}}
\def\bean{\begin{eqnarray*}}
\def\eean{\end{eqnarray*}}

\begin{document}

%\noindent Draft, June 2013
%
%\vspace{8mm}
% 
%%\begin{center}
%
%{\Large TRAPPED SURFACES IN}
%
%\vspace{10mm}
%
%{\Large  OPPENHEIMER-SNYDER BLACK HOLES}
%
%
%\vspace{10mm}
%
\title{Trapped Surfaces in Oppenheimer-Snyder Black Holes}

\author{Ingemar Bengtsson$^{1}$, Emma Jakobsson$^{1}$ and Jos\'e M. M. Senovilla$^{2}$\\ \\
$^{1}$ Stockholms Universitet, Fysikum\\
AlbaNova, S-106 91 Stockholm, Sweden\\ \\
$^{2}$ F\'isica Te\'orica, Universidad del Pa\'is Vasco\\
Apartado 644, 48080 Bilbao, Spain\\
\\
\footnotesize{e-mail addresses: ibeng@fysik.su.se,\, emma.jakobsson@fysik.su.se,\, josemm.senovilla@ehu.es}}

\date{}
\maketitle

\begin{abstract}
\noindent The Oppenheimer-Snyder solution models a homogeneous round dust cloud collapsing to a black hole. Inside its event horizon there is a region through 
which trapped surfaces pass. We try to determine exactly where the boundary 
of this region meets the centre of the cloud. We present explicit examples of the relevant trapped (topological) spheres; they
extend into the exterior vacuum region, and are carefully matched at the 
junction between the cloud and the vacuum.
\end{abstract}


%\newpage

%%%%%%%%%%%%%%%%%%%%%%%%%%%%%%%%%%%%%%%%%%%%%%%%%%%%%%%%%%%%%%%%%%%%%%%%%
%INTRODUCTION
%%%%%%%%%%%%%%%%%%%%%%%%%%%%%%%%%%%%%%%%%%%%%%%%%%%%%%%%%%%%%%%%%%%%%%%%%

\section{Introduction}
In spacetime terms the boundary of a black hole is -- according to a definition 
which may need refinement \cite{Hayward, Ashtekar, Booth} -- its 
event horizon. 
According to Penrose's singularity theorem \cite{Penrose} it is the 
appearance of trapped surfaces that really spells the doom of the collapsing 
matter. The event horizon is added as an afterthought by a cosmic censor. 
Indeed, in numerical relativity, the signal for a black hole is the presence of 
outer trapped surfaces on a given spatial slice \cite{Baumgarte}. In a 
dynamical situation these typically lie well inside the event horizon, 
but by considering all possible slicings outer trapped surfaces can probably 
be found passing through every point inside the event horizon \cite{Eardley}, 
while trapped surfaces cannot \cite{Bendov}. The distinction between trapped 
and outer trapped surfaces comes about because 
the latter are required to be (weakly) trapped, that is, to have (non-positive) negative future null expansions both 
outwards and inwards, thus obviating the need for using a spatial hypersurface 
to provide the meaning of outer. In this paper we are concerned with (weakly)
trapped surfaces only. 
 
The boundary of the region where trapped surfaces occur \cite{Hayward94} is remarkably 
difficult to determine \cite{BSlang}. This is true also for the simplest possible models 
of matter collapsing to form black holes, the Oppenheimer-Snyder (OS) and Vaidya 
solutions. Both of them are spherically symmetric, and are constructed by 
matching regions with collapsing matter to vacuum regions. They both have a 
central world line surrounded by a tube of round marginally trapped surfaces 
(MTS). In the Vaidya 
model this tube is spacelike, is composed of outermost stable MTS (in a 
technical sense \cite{Lars}), and lies outside 
the causal past of the central world line. In the OS model 
the tube is timelike, is composed of unstable MTS, and is visible in its 
entirety from the central world line. In some ways therefore the two models 
behave very differently. In the inhomogeneous Lema\^{i}tre-Tolman-Bondi 
models both kinds of behaviour are observed \cite{Brits}. An achronal spherically symmetric tube of MTS 
asymptotic to the event horizon will exist provided certain conditions 
on the stress-energy tensor are met \cite{Williams}.

Trapped surfaces are compact without boundary, in particular we will consider topological spheres, but they do not have to be round. 
So we can still ask whether there are trapped surfaces intersected by 
the central world line. For the Vaidya model, with some conditions on the 
rate of infall of matter, the answer is yes \cite{BSkort}, even though in this case the 
central world line never encounters non-zero spacetime curvature. 
Here we address the same question 
for the OS model. Since the answer to the first question is 
again yes, we go on to ask at what value of proper time along the central 
world line trapped surfaces are first encountered. We believe that we know the 
answer to this question too, but will not be able to offer a conclusive proof. 
At least, we have taken a step towards determining the location of the boundary 
of the trapped region in this model.  

In the construction of the models the matching of the matter filled regions to 
the vacuum regions is done in such a way that the spacetime metric is $C^1$, 
but this is not 
manifest in the coordinate descriptions used. We will insist that the trapped 
surfaces we consider have the same degree of differentiability. In the earlier 
paper on the Vaidya model \cite{BSkort} this issue was not properly addressed, but then 
the question was not very critical either because no attempt to optimize the 
construction was made there. For our purposes here it is crucial to handle 
this issue with care, and we explain the rules in a separate Appendix.    

We begin the story in section 2 by describing the Oppenheimer-Snyder solution 
in some detail, and comparing it to the Vaidya solution. Section 3 
contains some preliminary discussion of trapped surfaces confined within the 
dust cloud. In section 4 we introduce the class of trapped surfaces 
that we believe are the best if one wants them to reach the centre of the cloud 
at the earliest possible times. We also calculate what we believe to be the 
earliest possible time. This is a main result of our paper. Some of the trapped surfaces are built explicitly and we want to remark that they are probably difficult to be found in numerical approaches. In section 5 we 
present partial proofs that the results of section 4 are indeed optimal, but our 
surfaces must eventually enter the vacuum exterior in order 
to close, and the complications there are such that a full proof escapes us. 
Section 6 contains further discussion about trapped 
surfaces in the OS spacetime, and asks some questions we would like to see answered. Section 7 gives our conclusions.

%%%%%%%%%%%%%%%%%%%%%%%%%%%%%%%%%%%%%%%%%%%%%%%%%%%%%%%%%%%%%%%%%%%%%%%%%%
% THE OPPENHEIMER-SNYDER SOLUTION
%%%%%%%%%%%%%%%%%%%%%%%%%%%%%%%%%%%%%%%%%%%%%%%%%%%%%%%%%%%%%%%%%%%%%%%%%%

\section{The Oppenheimer-Snyder solution}
The Oppenheimer-Snyder solution consists of a piece of a $k = 1$ dust 
filled Friedmann model, matched across comoving spheres to a timelike 
hypersurface in the Schwarzschild solution ruled by timelike geodesics \cite{OS}. 
This solution was a midwife for the notion of black holes, and still plays an 
important role say as a model example for numerical relativity \cite{Staley}. 
Technicalities apart it is best explained by a picture. See Fig. 1, whose caption 
provides a reminder of the salient facts.

The technicalities are important for our purposes though. The metric within the 
dust cloud is 

\begin{eqnarray} ds^2 = - d\tau^2 + a^2(\tau )\left[ d\chi^2 + 
\sin^2{\chi}(d\theta^2 + \sin^2{\theta}d\phi^2)\right] = \nonumber \\ \ \\
= a^2(\eta)\left[ - d\eta^2 + d\chi^2 + \sin^2{\chi}(d\theta^2 + 
\sin^2{\theta}d\phi^2)\right] \ . \nonumber \end{eqnarray}

\noindent This is a solution of Einstein's equations if

\begin{equation} a(\eta) = \frac{a_{\rm m}}{2}(1-\cos{\eta}), \label{a} \end{equation}

\begin{equation} \tau (\eta) = \frac{a_{\rm m}}{2}(\eta - \sin{\eta}) \ , 
\label{tau} \end{equation} 
where $a_m$ is a constant determining the minimum energy density of the dust cloud. The dust is moving along timelike geodesics at constant $\chi$, 
$\theta$, $\phi$. We assume that we are in the collapsing phase ($\pi < \eta < 2\pi$, 
$a_{,\tau} < 0$), and moreover less than half of the 3-sphere is included 
($\chi \leq \chi_0 < \pi/2$) because we are going to match this solution to 
Schwarzschild at the comoving hypersurface $\chi = \chi_0<\pi/2$.

\begin{figure}[!t]
	\begin{center}
	\leavevmode
	\includegraphics[width=1.0\textwidth]{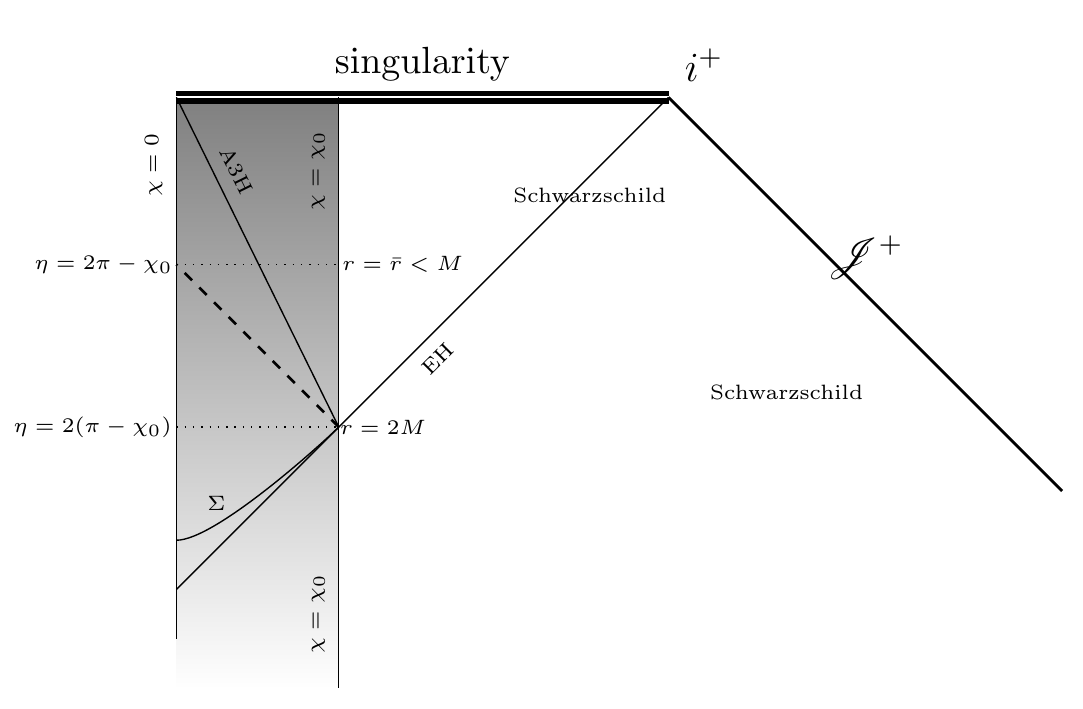}
	\end{center}
	\caption{\small A Penrose diagram of the OS black hole. The shadowed 
        region is a 
collapsing part of a closed Friedmann dust model, and the rest is Schwarzschild. 
The two regions are matched at a comoving timelike hypersurface 
$\chi = \chi_0$. (The coordinates used are explained in the text.) 
EH denotes the event horizon, and A3H denotes a timelike tube 
of marginally trapped surfaces. They meet in a sphere 
at the junction. The simple argument in section 3 shows 
that there are round trapped surfaces above the past light cone defined 
by the dashed line. Together with EH it defines a causal diamond in which 
the boundary of the trapped region must lie. The boundary must 
lie above the hypersurface $\Sigma$, which has constant Kodama time and 
forms a past barrier for trapped surfaces \cite{BSlang}. The interior time at the upper vertex of the diamond defines a round 2-sphere on the matching hypersurface whose area coordinate in the exterior is given by $r=\bar r = (M/2)\cos^{-2}(\chi_{0}/2)$, and this is always less than $M$, also indicated in the figure. }
	\label{fig:OS1}
\end{figure}

In the exterior we have the Schwarzschild metric 

\begin{equation} ds^2 = - V(r)dt^2 + \frac{dr^2}{V(r)} + r^2(d\theta^2 + 
\sin^2{\theta}d\phi^2) \ , \hspace{8mm} V(r) = 1 - \frac{2M}{r} \ . \end{equation}

\noindent We assume that $r \geq a \sin{\chi_0}$ in order to match 
the two solutions across the hypersurface 

\begin{equation} r = R(\eta ) = a\sin{\chi_0} \ . \label{R}\end{equation}

\noindent Eqs. (\ref{a},\ref{tau}) are still in force, and imply that 
this hypersurface is ruled by radially infalling Schwarzschild geodesics. 

The matching is done in such a way that the first and second fundamental forms 
of the hypersurface agree, from whatever side they are evaluated. The point 
of this requirement is to guarantee that there exists a coordinate system 
(not the ones we are using!) in which the metric is $C^1$ everywhere. The 
calculation is well explained in textbooks \cite{Poisson}. It is seen to 
relate the parameters of the solutions by 

\begin{equation} M = \frac{a_{\rm m}}{2}\sin^3{\chi_0} \ . \label{M} \end{equation}

\noindent This means that the Schwarzschild mass equals the Misner-Sharp mass 
of the Friedmann model evaluated at the junction hypersurface. Furthermore

\begin{equation} t = T(\eta ) \ , \hspace{8mm} T_{,\eta} = \frac{a\cos{\chi_0}}
{V(R)} \ . \label{T} \end{equation}

\noindent Due to translation invariance in the Schwarzschild part $t$ is 
determined only up to a constant.

As explained in the Appendix, and for the calculations we are going to perform, it is necessary to identify properly the tangent spaces at both sides of the matching hypersurface. First, the unit normal to the matching hypersurface has to be identified with the proper orientation, and this is done simply as 
\begin{equation} \label{m} 
n^{-}= ad\chi \quad \stackrel{identify}{\longleftrightarrow} \quad n^{+}= \frac{1}{a}(T_{,\eta}dr - R_{,\eta}dt) 
\end{equation}
at $\chi =\chi_{0}$, where (\ref{R}-\ref{T}) have been used. Then, the tangent vectors have to be identified properly. The angular part is identified in a natural way. Concerning the third tangent vector, we note that there is a uniquely defined timelike unit vector tangent to the matching hypersurface and orthogonal to 
the round spheres on both sides, and they are naturally identified,

\begin{equation} \label{s} e^{-}_{\eta}=\frac{1}{a}\partial _\eta \quad \stackrel{identify}{\longleftrightarrow} \quad  \frac{1}{a}\left( T_{,\eta}\partial_{t}+ R_{,\eta}\partial_{r}
\right),  \end{equation}
at $\chi=\chi_{0}$ where again (\ref{R}-\ref{T}) must be used.

Finally the tube of marginally trapped round spheres that we mentioned 
in the introduction is located at

\begin{equation} \label{mtt} \mbox{A3H}:\, a_{,\tau}^2 = \cot^2{\chi} \hspace{5mm} \Rightarrow \hspace{5mm} \eta = 2\pi - 2\chi
 \ . \end{equation}

\noindent Space is collapsing so quickly that any round sphere larger than this 
is trapped. 

There is a past barrier $\Sigma$ for trapped surfaces \cite{BSlang} defined by the concrete value of ``Kodama time'' such that the hypersurface $\Sigma$ meets the event horizon EH at  the matching hypersurface $\chi=\chi_0$, which happens at $\eta = 2(\pi -\chi_0)$. In the region below A3H (for $\eta<2(\pi -\chi)$) of the interior Friedmann part, Kodama time is given by constant values of $\cos^2\frac{\eta}{2}\cos\chi$,
and thus
$$
\Sigma : \hspace{5mm} \cos^2\frac{\eta}{2}\cos\chi =\cos^3\chi_0 \, .
$$

\begin{figure}[!t]
	\begin{center}
	\leavevmode
	\includegraphics[width=0.6\textwidth]{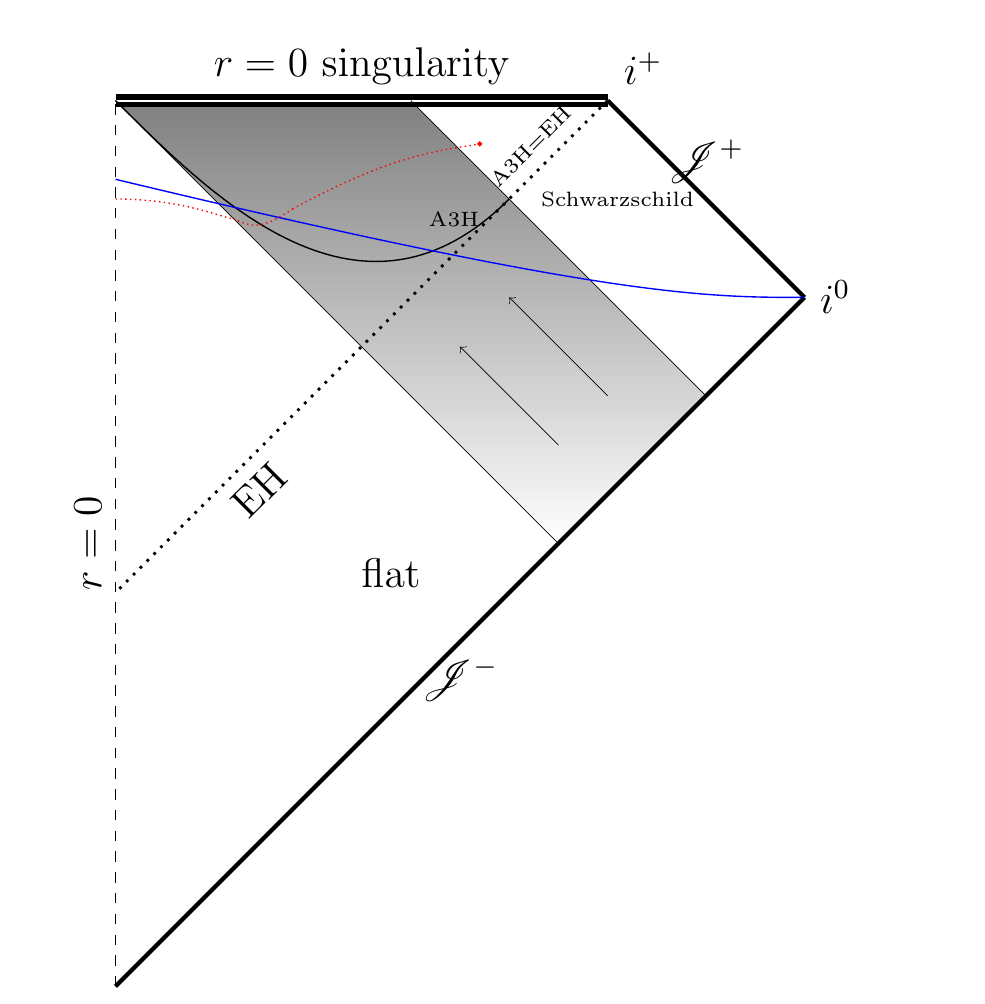}
	\end{center}
	\caption{\small A Penrose diagram of a Vaidya solution. The matter 
        region is filled with null dust. Again there is a tube of marginally round trapped 
        spheres within the matter region, but it is a spacelike tube of spheres 
        that cease to be trapped if deformed outwards within suitable hypersurfaces. For instance, the spacelike 
        hypersurface marked in blue intersects A3H twice, the inner intersection is unstable, while the outer is stable within the given hypersurface. It can be checked that outermost intersections are always stable in this sense. A (non-round) trapped surface going through the centre \cite{BSkort} is marked with red.}
	\label{fig:trap4}
\end{figure}

It is interesting to compare the OS solution to the Vaidya model, see Fig. \ref{fig:trap4}. The latter also has 
a tube of marginally trapped round spheres, but the Vaidya tube is spacelike, and 
its marginally trapped spheres are outermost stable, in the sense that if they are deformed outwards within a suitable spacelike hypersurface they cease to be trapped.
In the OS model, on the contrary, they cease to be trapped if deformed inwards. 
In this sense they are unstable \cite{Lars}. Several studies of non-spherically 
symmetric trapped surfaces in the Vaidya spacetime are available 
\cite{Bendov, BSkort, Schnetter}.

%%%%%%%%%%%%%%%%%%%%%%%%%%%%%%%%%%%%%%%%%%%%%%%%%%%%%%%%%%%%%%%%%%%%%%%%
% TRAPPED SURFACES WITHIN THE DUST CLOUD
%%%%%%%%%%%%%%%%%%%%%%%%%%%%%%%%%%%%%%%%%%%%%%%%%%%%%%%%%%%%%%%%%%%%%%%%

\section{Trapped surfaces within the dust cloud}
Before considering surfaces extending into both the Friedmann and the Schwarz\-schild region of the OS solution, let us warm up with round spheres contained within the dust cloud. We know that there is a marginally trapped tube \eqref{mtt} consisting of round spheres centred at $\chi=0$. However, since space is homogeneous, a round sphere within the dust cloud centred at some other value of $\chi$ must also be marginally trapped. Thus the marginally trapped round spheres inside the cloud can freely be moved around while still being marginally trapped, as long as they do not extend into the exterior Schwarzschild region.

\begin{figure}[!t]
	\begin{center}
	\leavevmode
	\includegraphics[width=1.0\textwidth]{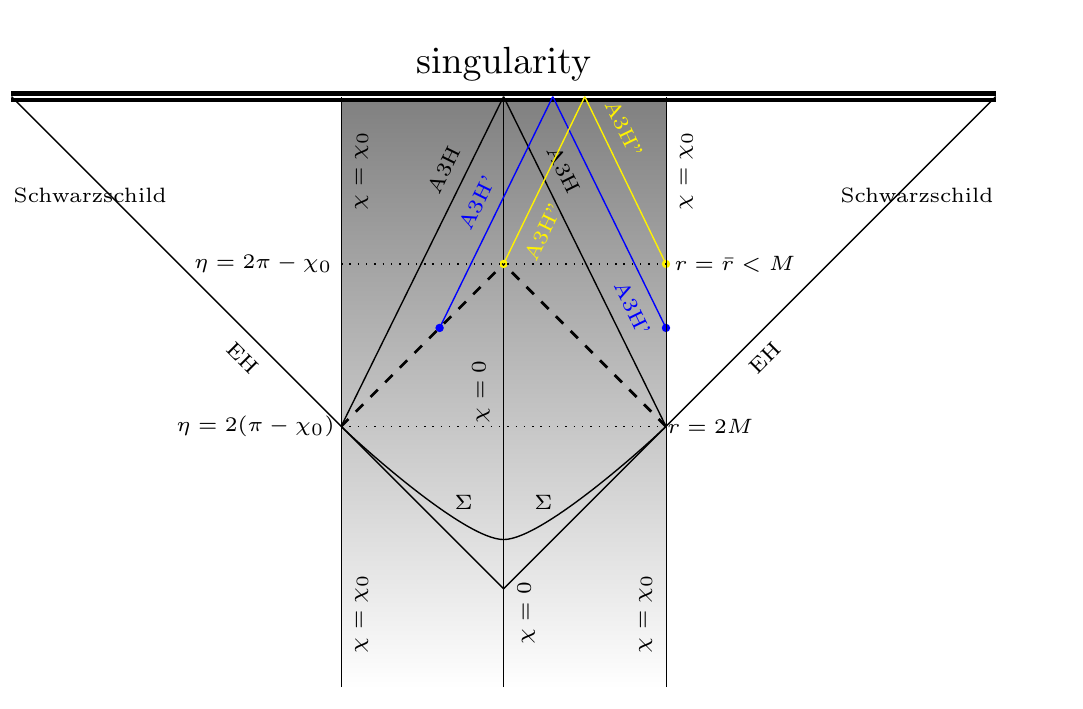}
	\end{center}
	\caption{\small A conformal diagram of the OS black hole. Round spheres centred at the origin are represented by two symmetrically placed points, so that the matching hypersurface here is represented by the two vertical lines with constant $\chi=\chi_0$. If the centred round spheres are situated above A3H they are trapped, and they are marginally trapped at A3H itself. However, as the shaded Friedmann region is spatially homogeneous -- so that the $\eta=$ constant hypersurfaces (horizontal lines in the shaded part of the diagram) are maximally symmetric -- we can move these round spheres, as well as A3H, and centre them anywhere on the slices as long as they do not enter the Schwarzschild region. This shows that there are trapped round spheres passing through every point of the interior for all $\eta>2\pi-\chi_0$. The dashed lines are obtained by shifting the marginally trapped tubes, centring them at values of $\chi\neq 0$, as for example the marginally trapped tube denoted A3H'. The two dots at the end represent a marginally trapped round sphere tangent to the matching hypersurface. The marginally trapped tube denoted A3H'' contains a marginally trapped round sphere -- represented by the dots -- tangent to both the matching hypersurface and the centre at $\eta=2\pi-\chi_0$.}
	\label{os2}
\end{figure}
In the conformal diagram of Fig. \ref{os2} the maneuver is illustrated by shifting the marginally trapped tube A3H to the left or right. Every pair of symmetrically placed points on the shifted tube represents a marginally trapped round sphere, as long as both points are contained within the dust cloud. Thus we immediately see that the region above the  dashed lines is filled with marginally trapped surfaces.

\begin{figure}[!t]
	\begin{center}
	\leavevmode
	\includegraphics[width=0.7\textwidth]{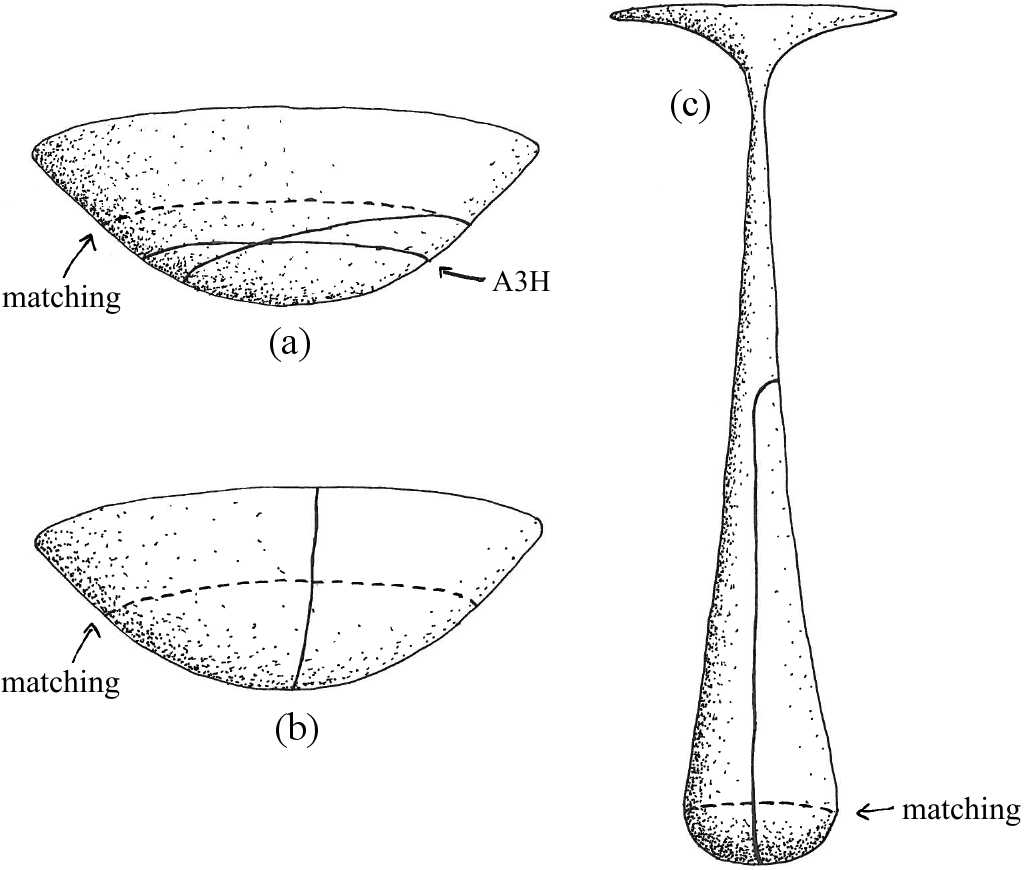}
	\end{center}
	\caption{\small Three different spatial slices of the OS model embedded in Euclidean space with one dimension suppressed. (a) A constant $\eta$ hypersurface in Friedmann matched to a spacelike hypersurface in Schwarzschild. Horizontal circles represent round spheres. The dashed circle is the sphere at $\chi=\chi_0$ where the matching is made. The continuous circles are marginally trapped spheres inside the dust cloud. One of them lies on A3H at a constant value of $\chi$. The other one is obtained by tilting the first one until it becomes tangent to the surface of the star. We see that the tilted circle reaches smaller values of $\chi$. (b) The same spatial slice as in (a) with the equatorial plane drawn as a black curve. This surface reaches the centre at $\chi=0$, but it also extends into the Schwarzschild region and it is not clear whether it can be closed or not. (c) The spatial slice on which the surface of Section \ref{first} lies. The surface is drawn as a black curve. At a small enough value of $r$ it deviates from the equatorial plane and is closed. Far from the cloud the hypersurface is bent so that it reaches spatial infinity -- rather than ending up in the singularity -- giving an intuitive definition of an ``outer'' direction on the surface.}
	\label{parad}
\end{figure}
We may also visualize the argument by drawing a picture of a spatial slice with $\eta$ constant inside the cloud. See Fig. \ref{parad}. Suppressing one dimension, the dust cloud -- which is a part of a 3-sphere -- can be drawn as part of a 2-sphere embedded in Euclidean space. Spheres of constant $\chi$ on the spatial slice are represented by horizontal circles on the spherical cap in the picture. The shifted spheres are illustrated in the picture by tilting these circles. By taking a circle representing a marginally trapped sphere and tilting it we find that it can reach smaller values of $\chi$ than the original one. But for values of $\eta$ smaller than $2\pi-\chi_0$ the centre will not be reached in this way, since that would require extending into the Schwarzschild region.

If we want to find a trapped surface passing through the centre at earliest possible time $\eta$ it must venture into the exterior. In Fig. \ref{parad} we see that we reach smaller values of $\chi$ the more we tilt the circles. The smaller the value of $\eta$ the more we need to tilt the marginally trapped surfaces in order to reach the centre. The intuitive strategy for optimizing the problem is thus to consider a circle tilted to the extreme so that it becomes vertical in the picture, i.e. to consider an equatorial plane.

%%%%%%%%%%%%%%%%%%%%%%%%%%%%%%%%%%%%%%%%%%%%%%%%%%%%%%%%%%%%%%%%%%%%%%%%%
% EQUATORIAL SURFACES PASSING THROUGH THE CENTRE
%%%%%%%%%%%%%%%%%%%%%%%%%%%%%%%%%%%%%%%%%%%%%%%%%%%%%%%%%%%%%%%%%%%%%%%%%

\section{Equatorial surfaces passing through the centre} \label{sec4}
We believe that surfaces confined to an equatorial plane have the best chance of reaching the centre of the dust cloud at early times, as argued in the previous section. Still we have to determine the exact shape of the surface in the interior as well as in the exterior, and the two pieces must be matched properly. Throughout we restrict ourselves to axially symmetric surfaces. We believe that allowing for more general surfaces will not improve the results, but we must admit that we do not have a proof of this statement.

The interior as well as the exterior is spherically symmetric, and despite the fact that we should have taken coordinates $\theta^{+},\phi^{+}$ in the interior and $\theta^{-},\phi^{-}$ in the exterior, by adapting them if necessary we can obviously drop the $\pm$ and take $\theta$ and $\phi$ as coordinates on the round spheres in the whole spacetime, and in particular on the matching hypersurface. Choosing the surface to lie in the equatorial plane $\theta=\pi/2$, we can describe it with local coordinates $\lambda$ and $\varphi$. As mentioned above, we assume that the surface is axially symmetric, and thus we set $\phi =\varphi$ so that on the interior part of the surface $\eta=\eta(\lambda)$ and $\chi=\chi(\lambda)$ are then functions only of $\lambda$, and in the exterior $r=r(\lambda)$ and $t=t(\lambda)$ are functions only of $\lambda$.

As explained in the Appendix there arise some constraint equations (\ref{xioflambda}) to ensure that the surface meets the matching hypersurface at the same set. The first of these constraints is
$$
\chi(\lambda)=\chi_{0} 
$$
and we assume that it has a solution given by $\lambda_{0}$. Then the value of $\eta$ at the intersection of the surface with the matching hypersurface is fixed and given by
$$
\eta_{0}\equiv \eta(\lambda_{0}) .
$$
The constraints involving the exterior part then become
$$
T(\eta_{0})=t(\lambda_{0})\equiv t_{0},Ê\quad \quad R(\eta_{0})=r(\lambda_{0})\equiv r_{0},
$$
where $t_{0}$ and $r_{0}$ denote the values of $t$ and $r$ at the intersection of the surface with the matching hypersurface. The first of these poses no problems due to the freedom in the choice of $T(\eta)$. Concerning the second, it leads to the basic relation
\begin{equation} \label{r0}
	r_0 = a(\eta_0)\sin\chi_0 \iff \cos\eta_0 = 1-\frac{r_0}{M}\sin^2\chi_0,
\end{equation}
determining the exterior value $r_0$ in terms of the interior values $\chi_0$ and $\eta_0$. It is also possible to think that it
determines the relation between $\eta_0$ and $r_0$ given the value of $\chi_0$.

The null normals to the surface, on which the expressions for the second fundamental forms depend, normalized such that ${k^\pm}^\mu k^{\pm}_{\mu} =-2$ are given by
\begin{equation} \label{kpm}
	{k^\pm}^\mu = \left\{
		\begin{array}{l l}
			\displaystyle{\frac{1}{a\sqrt{\chi'^2-\eta'^2}}(\chi'\partial_\eta+\eta'\partial_\chi) \pm \frac{1}{a\sin\chi}\partial_\theta} & \quad \text{on the Friedmann side,} \\
			\displaystyle{\frac{1}{\sqrt{\Delta}}\left(\frac{r'}{V}\partial_t+t'V\partial_r\right) \pm \frac{1}{r}\partial_\theta} & \quad \text{on the Schwarzschild side,}
		\end{array}
	\right.
\end{equation}
where the primes denote differentiation with respect to $\lambda$. 
With these null normals the null expansions are
\begin{equation} \label{theta}
	\theta^\pm = \left\{
		\begin{array}{l l}
			\displaystyle{\frac{1}{2a\sqrt{\chi'^2-\eta'^2}}\left(\frac{\chi'\eta''-\eta'\chi''}{\chi'^2-\eta'^2}+\eta'\cot\chi+2\chi'\cot\frac{\eta}{2}\right)} \\
			\displaystyle{-\frac{1}{2r\Delta^{3/2}}\left(r\left(r''-\frac{r'}{t'}t''\right)+\frac{M+r}{2M-r}r'^2+\frac{1}{r^2}(M-r)(2M-r)t'^2\right)}
		\end{array}
	\right.
\end{equation}
in the Friedmann and Schwarzschild parts, respectively.

The null normals (\ref{kpm}) agree on the matching hypersurface, in other words they comply with the necessary conditions (\ref{normals}) that can be computed using (\ref{m}) and (\ref{s}), if the following holds:
\begin{align}
	& \chi'(\lambda_0) = \left.\frac{1}{a^2}(r'T_{,\eta}-t'R_{,\eta})\right|_{\lambda_0}, \label{dchi} \\
	& \eta'(\lambda_0)=\left.\frac{1}{a^2V}(t'V^2T_{,\eta}-r'R_{,\eta})\right|_{\lambda_0}, \label{deta}
\end{align}
where (\ref{r0}) has been taken into account.
This fixes the first derivatives of $\chi(\lambda)$ and $\eta(\lambda)$ at $\lambda_{0}$ given those of $r(\lambda)$ and $t(\lambda)$ there. An equivalent version, interchanging both sides, reads
\begin{align}
	& r'(\lambda_0) = \left.\left(\chi'VT_{,\eta}+\eta'R_{,\eta}\right)\right|_{\lambda_0}, \label{dr} \\
	& t'(\lambda_0) = \frac{1}{V}\left.\left(\chi'R_{,\eta}+\eta'VT_{,\eta}\right)\right|_{\lambda_0}, \label{dt}
\end{align}
and can be obtained by solving \eqref{dchi} and \eqref{deta} as equations for $r'(\lambda_{0})$ and $t'(\lambda_{0})$.

We may now compute the first and second fundamental forms of the surface. The results are presented in Table \ref{table}. The $\Delta$ appearing in the table is defined as
\begin{equation} \label{Delta}
	\Delta = \frac{r'^2}{V}-t'^2V.
\end{equation}
It has to be positive for the surface to be spacelike. 

\renewcommand\arraystretch{2}
\begin{table}[!t] \footnotesize
	\centering
	\begin{tabular}{c|c|c|}
		\cline{2-3}
		& \textbf{Friedmann} & \textbf{Schwarzschild} \\ \hline
		\multicolumn{1}{|c|}{$\boldsymbol{\gamma_{\lambda\lambda}}$} & $a^2(\chi'^2-\eta'^2)$ & $\Delta$ \\ \hline
		\multicolumn{1}{|c|}{$\boldsymbol{\gamma_{\varphi\varphi}}$} & $a^2\sin^2\chi$ & $r^2$ \\ \hline
		\multicolumn{1}{|c|}{$\boldsymbol{\partial_\lambda\gamma_{\lambda\lambda}}$} & $2a^2\left(\chi'\chi''-\eta'\eta''+a_{,\tau}\eta'(\chi'^2-\eta'^2)\right)$ & $2\left(\frac{r'r''}{V}-t't''V-\frac{r'}{V}\frac{M}{r^2}\left(\frac{r'^2}{V}+t'^2V\right)\right)$ \\ \hline
		\multicolumn{1}{|c|}{$\boldsymbol{\partial_\lambda\gamma_{\varphi\varphi}}$} & $2a^2\sin^2\chi\left(a_{,\tau}\eta'+\chi'\cot\chi\right)$ & $2rr'$ \\ \hline
		\multicolumn{1}{|c|}{$\boldsymbol{K_{\lambda\lambda}}$} & $\frac{a}{\sqrt{\chi'^2-\eta'^2}}\left(\chi'\eta''-\eta'\chi''+a_{,\tau}\chi'(\chi'^2-\eta'^2)\right)$ & $\frac{1}{\sqrt{\Delta}}\left(r't''-t'r''+t'\frac{M}{r^2}\left(\frac{3r'^2}{V}-t'^2V\right)\right)$ \\ \hline
		\multicolumn{1}{|c|}{$\boldsymbol{K_{\varphi\varphi}}$} & $\frac{a\sin^2\chi}{\sqrt{\chi'^2-\eta'^2}}\left(a_{,\tau}\chi'+\eta'\cot\chi\right)$ & $\frac{1}{\sqrt{\Delta}}rt'V$ \\ \hline
	\end{tabular}
	\caption{\small The nonvanishing components of the first fundamental form $\gamma_{AB}$ and its derivatives and the second fundamental form $K_{AB}$ in the interior and the exterior for a surface in the equatorial plane. The primes indicate differentiation with respect to $\lambda$.} \label{table}
\end{table}
\renewcommand\arraystretch{1}

All the things listed in Table \ref{table} must be continuous across the matching hypersurface as explained in more detail in the Appendix. The continuity of the third fundamental form -- which is not listed in the table -- trivially holds since it vanishes on both sides of the matching hypersurface. Some of the continuity conditions of Table \ref{table} are already fulfilled due to (\ref{r0}) and either of (\ref{dchi},\ref{deta}) or (\ref{dr},\ref{dt}). The rest yields the remaining matching conditions for the surface, fixing the second derivatives as
\begin{align}
	& \chi''(\lambda_0) = \left.\left(\chi'\frac{\partial_\lambda\gamma_{\lambda\lambda}}{2\Delta}+\eta'\frac{K_{\lambda\lambda}}{\sqrt{\Delta}}-2\frac{R_{,\eta}}{r}\chi'\eta'\right)\right|_{\lambda_0}, \label{d2chi} \\
	& \eta''(\lambda_0) = \left.\left(\eta'\frac{\partial_\lambda\gamma_{\lambda\lambda}}{2\Delta}+\chi'\frac{K_{\lambda\lambda}}{\sqrt{\Delta}}-\frac{R_{,\eta}}{r}(\chi'^2+\eta'^2)\right)\right|_{\lambda_0}, \label{d2eta}
\end{align}
where the values of $\partial_\lambda\gamma_{\lambda\lambda}$ and $K_{\lambda\lambda}$ are the expressions given in Table \ref{table} on the Schwarzschild side evaluated at the matching hypersurface. As before, we can also write these conditions in an alternative way by interchanging the roles of both sides as
\begin{align}
	& r''(\lambda_0) = \left.\left(r'\frac{\partial_\lambda\gamma_{\lambda\lambda}}{2\Delta}+t'V\frac{K_{\lambda\lambda}}{\sqrt{\Delta}}+\frac{M}{r^2}\Delta\right)\right|_{\lambda_0}, \label{d2r} \\
	& t''(\lambda_0) = \left.\left(t'\frac{\partial_\lambda\gamma_{\lambda\lambda}}{2\Delta}+\frac{r'}{V}\frac{K_{\lambda\lambda}}{\sqrt{\Delta}}-\frac{2M}{r^2}\frac{r't'}{V}\right)\right|_{\lambda_0}, \label{d2t}
\end{align}
where $\partial_\lambda\gamma_{\lambda\lambda}$ and $K_{\lambda\lambda}$ are now evaluated at the matching hypersurface using the expressions on the Friedmann side.

In the rest of this section we will present two different choices for the functions of $\lambda$ describing the surface. In the first example the surface will be specified in Schwarzschild putting restrictions on the values of the functions $\eta(\lambda)$ and $\chi(\lambda)$ and their first and second derivatives at the junction. 
In our second example the surface will be specified in the Friedmann part, putting restrictions on how it can be continued into Schwarzschild.

% THE FIRST CONSTRUCTION

\subsection{The first construction} \label{first}
In this section we make a simple Ansatz for the surface and optimize that particular construction. This will then give us a hint of what the truly optimized solution might be.

We have already decided on sticking to the equatorial plane close to the centre of the cloud. However, the surface must eventually leave the equatorial plane in order to be closed. It has been shown \cite{BSkort} that an equatorial surface of constant $r$ within the Schwarzschild part can be closed properly -- that is, being closed while held trapped -- if the constant value of $r$ is small enough. Even though we want to keep things simple we have a better chance of succeeding in our quest if we add one degree of freedom letting not $r$ be constant but rather $dr/dt=r'/t'=k$ with $k$ constant. A small enough value of $r$ can always be reached if $k$ is negative. Once this value has been reached the surface can be bent into a surface of constant $r$ by letting $d^{2}r/dt^{2}= (r''-(r'/t')t'')/t'^{2}$ differ from zero in a small region. This is harmless, since we see from Eq. \eqref{theta} that a positive $r''-(r'/t')t''$ will only make the null expansions more negative.

Demanding that the surface is spacelike and that the null expansions are non-positive will put restrictions on the constant $k$. The surface is spacelike if the $\Delta$ of Eq. \eqref{Delta} is positive everywhere. We want to keep the surface inside the event horizon, so the function $V(r)$ is negative, and thus the condition becomes that $k>V$ everywhere. The function $V(r)$ obtains its maximum value when $r$ does, which by construction is at the junction. Thus the lower bound for $k$ is given by
\begin{equation} \label{lower}
	k > 1-\frac{2M}{r_0}.
\end{equation}
With our choice $r'(\lambda) =k t'(\lambda)$ the null expansions in Eq. \eqref{theta} are non-positive if 
\begin{equation} \label{k2}
	k^2 \geq \frac{r-M}{r+M}V^2.
\end{equation}
There are now three different cases. If $r_0$ -- and thus every other value of $r$ on the exterior part of the surface -- is less than $M$, the surface is safely trapped in the exterior. The right hand side of Eq. \eqref{k2} attains its maximum when
\begin{equation*}
	r = r_m = \frac{M}{3}(1+\sqrt{7})\approx 1.22M.
\end{equation*}
If $r_0$ is less than this value it is enough to check that the null expansions are non-positive at the junction, and if it is larger than this value it is enough to check that the null expansions are non-positive at $r=r_m$. We get the following upper bounds for $k$:
\begin{equation} \label{upper}
	k \leq \left\{
	\begin{array}{l l l}
		0 & \quad \text{if } r_0\leq M \\
		\left(1-\frac{2M}{r_0}\right)\sqrt{\frac{r_0-M}{r_0+M}} & \quad \text{if } M\leq r_0\leq r_m \\
		\left(1-\frac{2M}{r_m}\right)\sqrt{\frac{r_m-M}{r_m+M}} & \quad \text{if } r_0\geq r_m
	\end{array}
	\right.
\end{equation}
There is also an upper bound for $r_0$. When
\begin{equation}
	r_0 = 2M\left(1-\left(1-\frac{2M}{r_m}\right)\sqrt{\frac{r_m-M}{r_m+M}}\right)^{-1} =\frac{2M}{1+\sqrt{14\sqrt{7}-37}}\approx 1.66M, \label{upperlimit}
\end{equation}
the lower and the upper bounds for $k$ become equal, leaving only one choice for $k$. If $r_0$ is larger than this value the surface can not be both spacelike and trapped. Observe that there is plenty of room between the value $r=\bar r = \frac{M}{2\cos^{2}\frac{\chi_{0}}{2}}<M$ shown in Fig. \ref{fig:OS1} and this value where our surface can intersect the matching hypersurface.

The surface in the exterior is now specified by choosing a value of $r_0$ less than the value (\ref{upperlimit}), and a value of $k$ satisfying the conditions \eqref{lower} and \eqref{upper} given $r_0$. It must then be matched to a surface in the interior. Suppose that the equatorial surface in the interior is determined by a function $f$ such that $\eta=f(\chi)$. This function must satisfy
\begin{equation*}
	f(\chi_0) = \eta_0,
\end{equation*}
with $\eta_0$ given by the matching condition \eqref{r0}. The values of the first and second derivatives of $f$,
\begin{align*}
	& \frac{df}{d\chi} = \frac{\eta'(\lambda)}{\chi'(\lambda)}, \\
	& \frac{d^2f}{d\chi^2} = \frac{1}{\chi'^2(\lambda)}\left(\eta''(\lambda)-\chi''(\lambda)\frac{\eta'(\lambda)}{\chi'(\lambda)}\right),
\end{align*}
are determined at the junction by the matching conditions \eqref{dchi}, \eqref{deta}, \eqref{d2chi} and \eqref{d2eta}. If we make the simple Ansatz
\begin{equation} \label{Ansatz1}
	f(\chi) = k_0+k_1\chi^2+k_2\chi^4,
\end{equation}
the coefficients $k_0$, $k_1$ and $k_2$ are completely determined by the matching conditions.

\begin{figure}[!t]
	\begin{center}
	\leavevmode
	\includegraphics[width=0.7\textwidth]{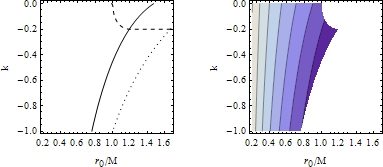}
	\end{center}
	\caption{\small The plots show how to choose $r_0$ and $k$ so that the value of $k_0$ becomes the smallest possible when $\chi_0=\pi/4$. The plot to the left helps us identify the allowed choices. Above the dotted curve the condition \eqref{lower} holds and the surface is spacelike. Below the dashed curve $k$ satisfies the inequality \eqref{upper}, with equality on the curve, and the surface is trapped in the exterior. To the left of the continuous curve the surface is trapped at the centre $\chi=0$, and on the curve it is marginally trapped there. We conclude that the allowed region can not extend past the continuous and dashed curves to the right. The contour plot to the right shows the value of $k_0$ with the forbidden (white) region excluded. The darker the shade, the smaller the value of $k_0$. We see that the best choice of $r_0$ and $k$ is the values they take where the continuous and the dashed curves intersect.}
	\label{kontur}
\end{figure}

The null expansions for the surface in the interior are given by Eq. \eqref{theta}. We want to choose $r_0$ and $k$ so that the coefficient $k_0$ of Eq. \eqref{Ansatz1} -- that is the value of $\eta$ at which the surface reaches the centre of the cloud -- is as small as possible while the null expansions are still non-positive. From Fig. \ref{kontur} we conclude that the best we can do is to choose the two constants $r_0$ and $k$ so that the surface is marginally trapped at the centre and so that it is either marginally trapped at the junction, if $r_0<r_m$, or marginally trapped at $r=r_m$, if $r_0>r_m$. The plots in Fig. \ref{kontur} are for $\chi_0=\pi/4$. But similar analyses for four other values of $\chi_0$ confirm the result. It may also be confirmed that with this choice of $r_0$ and $k$ the null expansions are negative for all values of $\chi$ between 0 and $\chi_0$ in the studied cases. The results for $k_0$ are shown in Table \ref{table2}. 
In Fig. \ref{ytor} the location of the best surface in a Penrose diagram of the Friedmann portion is shown for a few different values of $\chi_0$. A complete Penrose diagram, with the entire construction of the surface both in the interior and exterior, is presented in Fig. \ref{fig:OS5}. Yet another picture of the surface for $\chi_0=\pi/4$ -- here embedded in Euclidean space -- is shown in Fig. \ref{parad}.

%THE SECOND CONSTRUCTION

\subsection{The second construction} \label{2nd}
Restricted to the previous construction we found that the optimal choice of the surface was such that it was marginally trapped at some critical points. The result can be made slightly better by choosing the surface so that it is kept marginally trapped everywhere inside the dust cloud and then matching it to some suitable surface in the exterior.
\renewcommand\arraystretch{1.5}
\begin{table}[!ht]
	\centering
	\begin{tabular}{r|c||c|c|c|c|c|}
		\cline{2-7}
		& $\boldsymbol{\chi_0}$ & $\pi/12$ & $\pi/6$ & $\pi/4$ & $\pi/3$ & $5\pi/12$ \\ \cline{2-7}
		first construction & $\boldsymbol{\eta(0)}$ & 5.7693 & 5.2716 & 4.8037 & 4.3752 & 4.0081 \\ \cline{2-7}
		& $\boldsymbol{X}$ & 0.037005 & 0.067999 & 0.11628 & 0.17797 & 0.26197 \\ \cline{2-7}
		second construction & $\boldsymbol{\eta(0)}$ & 5.77 & 5.26 & 4.79 & 4.36 & 4.00 \\ \cline{2-7}
		& $\boldsymbol{X}$ & 0.040 & 0.046 & 0.099 & 0.16 & 0.26 \\ \cline{2-7}
	\end{tabular}
	\caption{\small The smallest values of $\eta$ at the centre for the two constructions at different values of $\chi_0$. The first construction gives an optimal solution and the result is given to five digits accuracy. In the second construction the values are the smallest possible to three digits accuracy so that the surface becomes well-behaved in the exterior. The smaller the dimensionless number $X$, the sooner the surfaces meet the centre compared to a given reference time.} \label{table2}
\end{table}
\renewcommand\arraystretch{1}

We want the surface to reach the centre $\chi=0$, and we also keep it in an equatorial plane with axial symmetry; hence it lies on a spacelike spherically symmetric hypersurface. There is a general result \cite{S} that implies that then the surface cannot be (marginally) trapped at any local maximum of $\chi$ in the region below A3H. Thus, we can safely choose $\chi(\lambda)=\lambda$ so that the null expansions \eqref{theta} in the interior become
\begin{equation} \label{theta2}
	\theta_\pm = \frac{1}{2a\sqrt{1-\eta'^2}}\left(\frac{\eta''}{1-\eta'^2}+\eta'\cot\chi+2\cot\frac{\eta}{2}\right).
\end{equation}
Putting these to zero gives a differential equation for $\eta(\lambda)$. If we give as initial conditions the value $\eta(0)$ at the centre and demand $\eta'(0)=0$ this differential equation can be solved numerically. This will completely determine the surface in the interior.

\begin{figure}[!t]
	\begin{center}
	\leavevmode
	\includegraphics[width=0.9\textwidth]{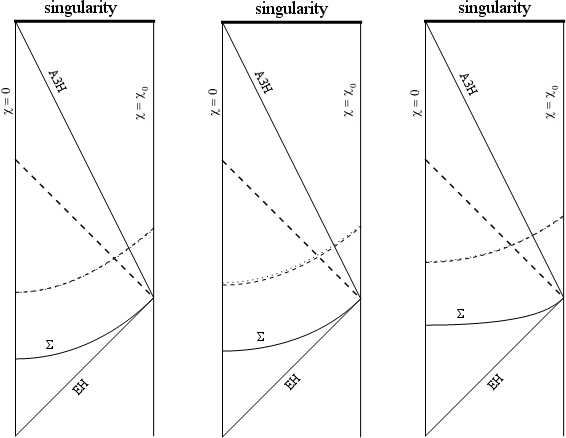}
	\end{center}
	\caption{\small Penrose diagrams of the interior of the dust cloud with $\chi_0=\pi/12$, $\chi_0=\pi/4$ and $\chi_0=5\pi/12$ from left to right. The dotted curves represent the best surfaces we got in the first construction. A point on each curve really represents a sphere in the Penrose diagram, but only the equator of each such sphere is part of the surface. The dashed curves represent the marginally trapped surfaces in the second construction. The two results are hardly separable to the eye in this figure.}
	\label{ytor}
\end{figure}

\begin{figure}[!ht]
\includegraphics[width=1.0\textwidth]{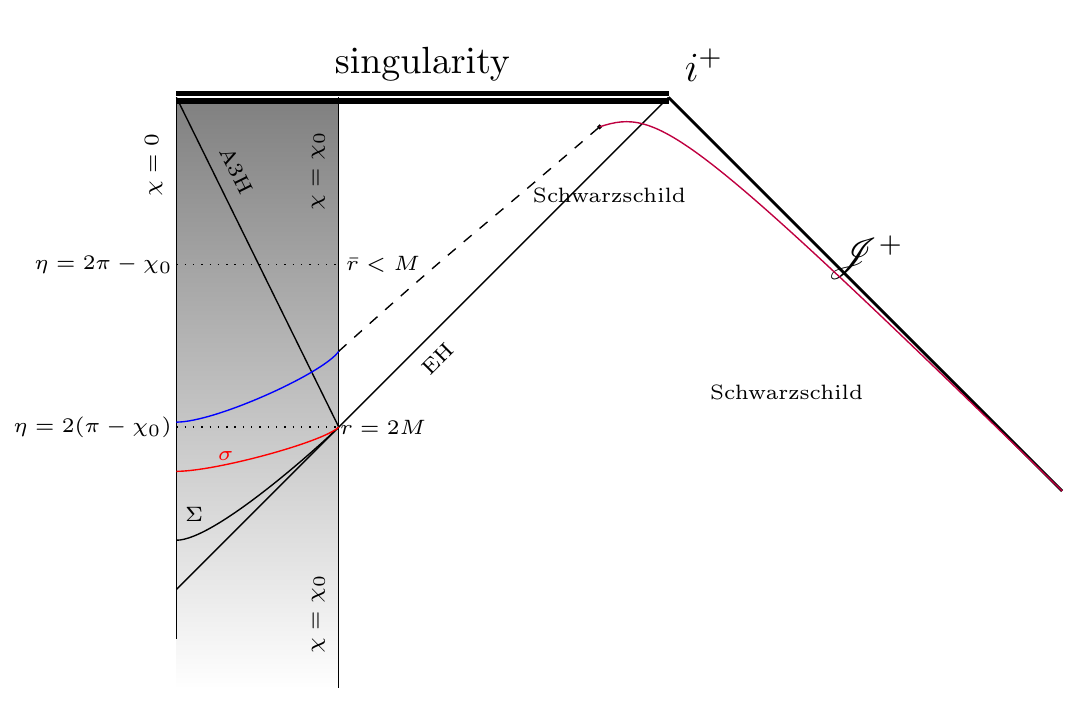}
\caption{\footnotesize{The OS black hole once again, with some novel decoration. The equatorial planes of the blue hypersurface can be joined to appropriate cylinders in the Schwarzschild region -- represented here by the dashed line -- such that they can eventually be capped -- the final dot -- to produce topological spheres that are (weakly) trapped , as has been explicitly demonstrated in the main text. The purple line together with the dashed and blue lines indicates how a possible spacelike and asymptotically flat hypersurface containing the trapped surface looks like in the diagram. This hypersurface corresponds to the spatial slice shown in Fig. \ref{parad}(c). The particular hypersurface called $\sigma$ shown in red has minimal equatorial planes and joins the past barrier $\Sigma$ and the EH at the round 2-sphere with $\chi =\chi_{0}$ and $\eta = 2(\pi -\chi_{0})$, as represented. In the main text we prove that $\sigma$ is a past barrier for axially symmetric trapped surfaces which are confined to equatorial planes within the interior part. One wonders if $\sigma$ can be promoted to a new past barrier for more general trapped surfaces, leaving even less room for the boundary $\B$ to be placed. }}
\label{fig:OS5}
\end{figure}

We must then make an Ansatz for the surface in the exterior. Suppose that the surface lies in the equatorial plane with $r=g(t)$ for some function $g$. Let us set the value of $t$ at the junction to zero: $t_{0}=t(\lambda_{0})=0$. Then we must have that $g(0)=r_0$ with the value of $r_0$ determined by the matching condition \eqref{r0}. The first and second derivatives of $g$ are
\begin{align*}
	& \frac{dg}{dt} = \frac{r'(\lambda)}{t'(\lambda)}, \\
	& \frac{d^2g}{dt^2} = \frac{1}{t'^2(\lambda)}\left(r''(\lambda)-t''(\lambda)\frac{r'(\lambda)}{t'(\lambda)}\right),
\end{align*}
and their values at $\lambda_{0} \Longleftrightarrow t=0$ are determined by the matching conditions \eqref{dr}, \eqref{dt}, \eqref{d2r} and \eqref{d2t}. A simple Ansatz for $g$ is to put
\begin{equation*}
	g(t) = g(0)+g'(0)t+\frac{1}{2}g''(0)t^2.
\end{equation*}
The whole construction is then completely determined by the value we choose for $\eta(0)$.

We want to choose $\eta(0)$ as small as possible but need to make sure that the chosen value makes the function $g(t)$ well-behaved in the exterior -- meaning that the surface can be closed and that the null expansions are negative. We do not give an explicit description of how the surface is to be closed in Schwarzschild, but we know that as long as small enough values of $r$ can be reached it can be done \cite{BSkort}. The values of $\eta(0)$ presented in Table \ref{table2} are such that the function $g(t)$ reaches values at least smaller than $0.02M$, and such that the null expansions are negative in the exterior. The result is drawn next to the surface resulting from the first construction in the Penrose diagrams of Fig. \ref{ytor} for three different values of $\chi_0$. It is striking how close the two results are to each other.

The value of $\eta(0)$ for the optimal trapped surface depends on the value of $\chi_{0}$ as we can check in Table \ref{table2}. To compare these different values, in order to see how the trapped surfaces extend to the centre depending on the physical parameters of the black hole, we take as reference time for all possible black holes $\eta =2(\pi-\chi_{0})$, that is, the first instant in $\eta$ with a marginally trapped round sphere. We want to compute how close to this reference time the trapped surfaces we have constructed can be. To that end, we define the dimensionless number
$$
X\equiv \frac{\eta(0)-2(\pi-\chi_{0})}{\chi_{0}} 
$$
and the smaller its value, the sooner the trapped surfaces are met at the centre of the dust cloud. These values are presented for both constructions in Table \ref{table2}. The second construction is better than the first in all cases except for the case $\chi_0=\pi/12$. For the latter case a more accurate analysis of the second construction would be needed to detect any difference. Not taking the smallest value of $\chi_0$ into consideration, we can check that $X$ is an increasing function of $\chi_{0}$. We draw the conclusion that the smaller the value of $\chi_{0}$, or in physical terms the smaller the mass $M$ of the black hole for a fixed energy density of the cloud, the sooner the trapped surfaces reach the centre of the cloud with respect to the reference time. Another way of putting this is: the comoving time $\eta$ elapsed between the first appearance of a marginally trapped round sphere and that of a trapped surface reaching the centre increases with the mass of the OS black hole (for a fixed energy density of the dust).

%%%%%%%%%%%%%%%%%%%%%%%%%%%%%%%%%%%%%%%%%%%%%%%%%%%%%%%%%%%%%%%%%%%%%%%%%%
% THE OPTIMAL CONSTRUCTION?
%%%%%%%%%%%%%%%%%%%%%%%%%%%%%%%%%%%%%%%%%%%%%%%%%%%%%%%%%%%%%%%%%%%%%%%%%%

\section{The optimal construction?}
In this section we discuss whether the construction in Section \ref{sec4} is optimal or not. There are some general results we can reach by considering only the Friedmann part of the surface, but still it must be admitted that the final result is dependent on the closing of the surface in Schwarzschild. The surfaces constructed in Section \ref{2nd} are minimal inside the dust cloud. We will here prove that any minimal surface in the equatorial plane within the dust cloud reaches values of $\eta$ at the centre that are lower than those reached by any other possible axially symmetric trapped surface crossing any round sphere intersected by the former in the interior region. We do this in three steps: first we consider minimal surfaces and show that, for MTS on equatorial planes, $\eta$ is an increasing function of $\chi$ everywhere in the dust cloud; second,  we prove that the claim holds for axially symmetric surfaces restricted to an equatorial plane and, finally, we remove the equatorial plane condition.

%MINIMAL EQUATORIAL PLANES

\subsection{Minimal equatorial planes}
Let us start by proving that any (piece of a) minimal surface can not have a local maximum of $\eta$ within the Friedmann part of the spacetime. For a two-dimensional surface $S$ with tangent vectors $\vec{e}_A$ and mean curvature vector $\vec{H}$ the following relation holds \cite{BSlang}:
\be \label{magic}
	\frac{1}{2}\gamma^{AB}e^\mu_Ae^\nu_B\left(\left.\pounds_{\vec{\xi}}g\right|_S\right)_{\mu\nu} = \overline{\nabla}_C\overline{\xi}^C+\xi_\alpha H^\alpha,
\ee
for any vector field $\vec{\xi}$. In Eq. \eqref{magic} $\overline{\nabla}_A$ is the Levi-Civita connection of the first fundamental form $\gamma_{AB}$, i.e. $\overline{\nabla}_A\gamma_{BC}=0$, and $\overline{\xi}_A$ is the projection of $\vec{\xi}$ on the surface, that is $\overline{\xi}_A=\left.\xi_\mu\right|_S e^\mu_A$. Let us choose $\vec{\xi}=\partial_\eta$, which is a future-pointing, timelike, conformal Killing vector field orthogonal to the spacelike hypersurfaces of constant $\eta$. Then
\ben
	\left(\pounds_{\vec{\xi}}g\right)_{\mu\nu} = 2a_{,\tau}g_{\mu\nu},
\een
and
\ben
	\overline{\xi}_A = \xi_\mu e^\mu_A = -a^2\overline{\nabla}_A\eta.
\een
If the surface is minimal $\vec{H}=\vec{0}$, and we find that with our choice of $\vec{\xi}$ Eq. \eqref{magic} becomes
\be \label{magic2}
	2a_{,\tau} = -a^2\left(\overline{\nabla}_A\overline{\nabla}^A\eta+2a_{,\tau}\overline{\nabla}_A\eta\overline{\nabla}^A\eta\right).
\ee
At a local extremum $q\in S$ of $\eta$ we have that $\overline{\nabla}_A\eta|_{q}=0$, so Eq. \eqref{magic2} becomes
\ben
	\left.\overline{\nabla}_A\overline{\nabla}^A\eta \right|_{q}=\left.-2\frac{a_{,\tau}}{a^2}\right|_{q} >0.
\een
The right-hand side of this equation is positive and thus any local extremum of $\eta$ must be a minimum. This applies, in particular, to axially symmetric marginally trapped surfaces confined to an equatorial plane because, as we know from (\ref{theta}), both expansions are equal and thus any MTS is actually minimal there.

There are also some conclusions we can draw on the behaviour of trapped and marginally trapped surfaces at the centre. Consider an axially symmetric surface in an equatorial plane and, as in Section \ref{2nd}, let us choose the parameter $\lambda$ on the surface so that $\chi(\lambda)=\lambda$, $0\leq\lambda\leq\chi_0$. The null expansions are then given by Eq. \eqref{theta2}. We see that any surface must have a local extremum at the centre:
\be \label{initial2}
	\eta'(0) = 0,
\ee
or otherwise the second term of Eq. \eqref{theta2} diverges. Then we have that
\ben
	\lim_{\lambda\to 0}\frac{\eta'(\lambda)}{\tan\lambda} = \lim_{\lambda\to 0}\frac{\eta''(\lambda)}{1+\tan^2\lambda} = \eta''(0),
\een
so that the null expansions at the centre are
\be\label{expcentre}
	\left.(a\theta_\pm)\right|_{\lambda=0} = 2\left(\eta''(0)+\cot\frac{\eta(0)}{2}\right).
\ee
They are non-positive if
\be \label{d2eta0}
	\eta''(0) \leq -\cot\frac{\eta(0)}{2} \quad (>0).
\ee
We see that for a trapped surface this local extremum can be either a minimum or a maximum. But for a marginally trapped (ergo minimal) surface the equality sign of Eq. \eqref{d2eta0} holds, and the surface must have a minimum at the centre.

From the results so far we can draw the conclusion that for an axially symmetric marginally trapped (ergo minimal) surface in an equatorial plane the function $\eta(\lambda)$ has a local minimum at the centre and -- since it can not have a maximum -- is everywhere an increasing function of $\lambda$. Since Eq. \eqref{initial2} holds for trapped as well as minimal surfaces at the centre, they are tangent there. Also we see from Eq. \eqref{d2eta0} that the value of the second derivative of $\eta(\lambda)$ at the centre for any trapped surface is smaller than the corresponding value for a minimal surface. Thus an axially symmetric trapped surface in an equatorial plane is, close to the centre, locally to the past of the minimal equatorial plane passing through the centre at the same instant of time.

%AXIALLY SYMMETRIC TRAPPED SURFACES ON EQUATORIAL PLANES

\subsection{Axially symmetric trapped surfaces on equatorial planes}
We will now give a definite proof that a minimal equatorial plane reaches the centre at an earlier time than any other axially symmetric trapped surface on an equatorial plane that happens to cross any round sphere touched by the minimal one. In Section \ref{2nd} we found the minimal equatorial planes by solving the differential equation obtained by putting Eq. \eqref{theta2} to zero, with initial conditions $\eta(0)=\alpha$, $\eta'(0)=0$. Let us denote the solution by $\eta_m(\lambda;\alpha)$. For each value of $\alpha$ there is a spacelike hypersurface $\eta=\eta_m(\lambda;\alpha)$ whose equatorial plane is a minimal surface, and the set of all these hypersurfaces defines a foliation of spacetime. We can define a time coordinate $T(\alpha)$ which is constant on each of these hypersurfaces. There is then a function $p(\lambda;\alpha)$ such that for each value of $T$ the hypersurface is defined by
\ben
	\eta-p(\lambda;\alpha) = T(\alpha).
\een
It is related to $\eta_m(\lambda;\alpha)$, also defining the foliation, by
\be \label{p}
	p(\lambda;\alpha)=\eta_m(\lambda;\alpha)-T(\alpha).
\ee
We do not have an explicit expression neither for $\eta_m(\lambda;\alpha)$ nor $T(\alpha)$, and hence not for $p(\lambda;\alpha)$. However, this will not be needed for the proof.
Note though that Eq. \eqref{p} implies that
\be \label{derivator}
	p'(\lambda;\alpha) = \eta_m'(\lambda;\alpha).
\ee
A vector field normal to the hypersurfaces is proportional to
\be \label{v}
	\vec{v} = \partial_\eta+p'\partial_\chi.
\ee
Our purpose is to show that any axially symmetric trapped surface in an equatorial plane can not have a minimum in $T$. Then, it is clear that such a trapped surface must reach the centre at a later time $T$ -- that is at a larger value of $\eta$ -- than any minimal equatorial plane it crosses, since minimal surfaces lie on hypersurfaces of constant $T$.

The proof is as follows: Consider a surface in the equatorial plane having a minimum in $T$ at a point $q$. Since the future-directed vector $\vec{v}$ is orthogonal to hypersurfaces of constant $T$ we have $v_a = -h\nabla_aT$ for some function $h>0$, or projected to the surface
\be \label{v_A}
	\bar{v}_A = -h\overline{\nabla}_AT.
\ee
If the surface has a local extremum of $T$ at $q$, then $\vec{v}$ is normal to the surface at that point. Thus
\be \label{v_Aq}
	\bar{v}_A|_q = 0,
	\ee
implying that $\overline{\nabla}_AT|_q=0$. Then from Eq. \eqref{v_A} we see that the divergence of $\bar{v}_A$ at $q$ becomes
\ben
	\left.\overline{\nabla}^A\bar{v}_A\right|_q = \left.-h\overline{\nabla}^A\overline{\nabla}_AT\right|_q.
\een
If the surface has a minimum of $T$ at $q$, then $\overline{\nabla}^A\overline{\nabla}_AT|_q>0$, implying that
\be \label{divv}
	\left.\overline{\nabla}^A\bar{v}_A\right|_q<0.
\ee
We now show that for an axially symmetric surface $S$ in the equatorial plane Eq. \eqref{divv} is not consistent with the surface being trapped. On $S$ set $\chi(\lambda)=\lambda$ so that the first fundamental form of the surface is
\ben
	\gamma_{AB}d\lambda^Ad\lambda^B = a^2\left[(1-\eta'^2)d\lambda^2+\sin^2\lambda d\varphi^2\right],
\een
and it has a timelike normal vector orthogonal to the round spheres of constant $\eta$ and $\chi$ given by $\partial_\eta+\eta'\partial_\chi$. If $S$ has a minimum of $T$ at the point $q$ the vector \eqref{v} is normal to the surface at that point implying that
\be \label{derivator2}
	\eta'(\lambda)|_q = p'(\lambda;\alpha_0)|_q,
\ee
with $\alpha_0$ given by
\be \label{etaq}
	\eta(\lambda)|_q=\eta_m(\lambda;\alpha_0)|_q.
\ee
The projection of $\vec{v}$ to the surface is
\ben
	\bar{v}_A = a^2|_S\left(-\overline{\nabla}_A \eta(\lambda)+p'(\lambda;\alpha)\overline{\nabla}_A\chi(\lambda)\right).
\een
Since Eq. \eqref{v_Aq} holds, the divergence of $\bar{v}_A$ becomes
\begin{align*}
	\left.\overline{\nabla}^A\bar{v}_A\right|_q & = a^2\left.\left(-\overline{\nabla}^A\overline{\nabla}_A\eta(\lambda)+p'(\lambda;\alpha_0)\overline{\nabla}^A\overline{\nabla}_A\chi(\lambda)+p''(\lambda;\alpha_0)\gamma^{\lambda\lambda}\right)\right|_q \\
	& = \left.\frac{p''(\lambda;\alpha_0)-\eta''(\lambda)}{1-\eta'^2(\lambda)}\right|_q,
\end{align*}
where Eq. \eqref{derivator2} has been used in the last step. With the point $q$ being a local minimum of $T$, this quantity must be negative according to \eqref{divv}, so that
\be \label{olikhet}
	p''(\lambda;\alpha_0)|_q < \eta''(\lambda)|_q.
\ee

The null expansions of the surface are given by Eq. \eqref{theta2}. Using \eqref{derivator}, \eqref{derivator2}, \eqref{etaq}, and the definition of the function $\eta_m$, we find that
\ben
	\theta_\pm|_q = \left.\frac{\eta''-p''}{2a(1-\eta_m'^2)^{3/2}}\right|_q.
\een
But if the inequality \eqref{olikhet} holds, the null expansions are positive and the surface is not trapped at $q$.

Thus we draw the conclusion that if we only consider axially symmetric surfaces in the equatorial plane, then minimal surfaces reach the centre at earliest possible times. This result supports the statement that the surfaces of Section \ref{sec4} are the best we can do.

%GENERAL AXIALLY SYMMETRIC TRAPPED SURFACES

\subsection{General axially symmetric trapped surfaces}
Consider finally any axially symmetric surface $S$ within the dust cloud. They can be described by the following parametric expressions
$$
\eta=\eta(\lambda), \quad \chi =\chi(\lambda), \quad \theta =\theta(\lambda), \quad \phi = \varphi
$$
where $\{\lambda^{B}\}=\{\lambda,\varphi\}$ are local coordinates on $S$. The tangent vectors are
$$
\vec e_{\lambda}= \eta' \partial_{\eta}+ \chi' \partial_{\chi}+\theta' \partial_{\theta},Ê\quad \quad \vec e_{\varphi}=\partial_{\phi} \, .
$$
 The first fundamental form reads
\be
\gamma_{AB}d\lambda^{A}d\lambda^{B}=a^{2}\left(\Delta_{F}^{2}d\lambda^{2}+\sin^{2}\chi(\lambda) \sin^{2}\theta(\lambda)  d\varphi^{2}\right)
\ee
with $\Delta_{F}^{2}= \chi'^{2}-\eta'^{2}+\theta'^{2}\sin^{2}\chi(\lambda) >0$. The future-pointing null normals can be chosen to be
$$
k^{\pm}_{\mu}dx^{\mu}=\frac{a}{\Delta_{F}}\left[-\delta^2 d\eta +\left(\eta'\chi'\pmÊ\theta'\Delta_F \sin\chi \right)d\chi +\left(\eta'\theta'\sin\chi \mp \chi'\Delta_F \right)\sin\chi d\theta\right]
$$
and they satisfy ${k^+}^\mu k^-_\mu =-2\delta^2$ with
$$
\delta^2 = \chi'^2+\theta'^2\sin^2\chi(\lambda) >0 \, .
$$
Then, a somewhat lengthy but straightforward calculation produces the null expansions
\bea
\theta^\pm = \frac{1}{a\Delta_F}\left\{2\delta^2\cot\frac{\eta}{2} +\frac{1}{\Delta^2_F}\left[\delta^2 \eta'' -\eta'\chi'\chi'' +(\chi'^2-\eta'^2) \chi'\eta'\cot\chi -\eta'\theta'\theta''\sin^2\chi\right] \right. \nonumber\\
\left. +\eta'\theta' \cot\theta \pm \frac{1}{\Delta_F} \left[\sin\chi \left(\chi'\theta''-\theta'\chi''Ê\right)+\theta'\cos\chi \left(\Delta_F^2+\delta^2+\chi'^2 \right)Ê-\Delta_F^2\chi'\frac{\cot\theta}{\sin \chi}\right]\right\}.\label{exp}
\eea
Assume that $S$ reaches the centre $\chi =0$, and set $\chi(0)=0$, i.e. $\lambda =0$ at the centre. The last summand in (\ref{exp}) remains finite only if $\cot \theta(0) =0$ implying that, at the centre, $\theta(0)=\pi/2$. (Observe that the term with $\cot \chi (0)$ cannot compensate for any divergence in $\cot \theta(0)/\sin\chi(0)$ for {\em both signs}).  The other critical term at the centre behaves like $
\chi'(0) \eta'(0) \cot \chi(0)$ and this necessarily requires that $\eta'(0)=0$ ($\chi'(0)=0$ is not possible as this would make the divergences even worse and would also violate the condition $\Delta_F^2 >0$). 

All in all, we can set around the centre $\chi =\lambda$ and the null expansions (\ref{exp}) become at the centre, by taking the appropriate limits,
$$
\theta^\pm (0) =\frac{1}{a} \left(2\eta''(0) +2\cot\frac{\eta(0)}{2}\pm 4\theta'(0)  \right),
$$ 
and this can be rewritten as
$$
\theta^\pm (0) = \Theta \pm\frac{4}{a}\theta'(0),
$$
where $\Theta$ is the value (\ref{expcentre}) of the null expansions at the centre for another surface $\tilde S$ on an equatorial plane {\em with the same} $\eta(\lambda)$. If $S$ is (marginally) trapped at the centre $\theta^\pm (0) \leq 0$ for both signs, and thus it is necessary that $\Theta \leq 0$ -- equality can only occur if $\theta'(0)=0$. This means that $\tilde S$ is (marginally) trapped at the centre, hence we know from the result in the previous subsection that $\tilde S$ reaches the centre at a later or equal time (i.e. at larger or equal values of $\eta$) than the minimal equatorial plane passing through the same round sphere as $\tilde S$. But then so does $S$, which has the same $\eta (\lambda)$. 

Thus, we have proven that general axially symmetric surfaces within the interior region reach the centre at later times than the equatorial minimal planes. Of course, the entire discussion has been restricted to the interior Friedmann region, and thus one can still wonder if there exist alternative strategies leading to better values of $\eta(0)$. 
Recall that what prevented us from pushing the surfaces of Section \ref{2nd} further down was that we had to make sure that they could be closed properly. But we do not know any optimal strategy of closing the surfaces, and are therefore unable to prove that they can not be pushed even further down.

A consequence of the results of this section is that we can define a past barrier for axially symmetric trapped surfaces reaching the centre. It is the hypersurface of constant $T$ that meets the event horizon at the junction between Friedmann and Schwarzschild, as shown in Fig.\ref{fig:OS5}. We call this hypersurface $\sigma$ in analogy with the past barrier $\Sigma$ of Fig. \ref{fig:OS1}. However, we do not know if $\sigma$ really is a past barrier for arbitrary trapped surfaces, while $\Sigma$ certainly is.

%%%%%%%%%%%%%%%%%%%%%%%%%%%%%%%%%%%%%%%%%%%%%%%%%%%%%%%%%%%%%%%%%%%%%%%%%%
% DISCUSSION
%%%%%%%%%%%%%%%%%%%%%%%%%%%%%%%%%%%%%%%%%%%%%%%%%%%%%%%%%%%%%%%%%%%%%%%%%%

\section{Discussion}
In the OS model the boundary $\B$ of the trapped region must be a spherically 
symmetric hypersurface \cite{BSlang} meeting the event horizon at the junction between 
dust and vacuum. We believe that the results of the previous sections are enough to 
pinpoint where $\B$ meets the central world line, and they are certainly 
consistent with $\B$ being spacelike throughout the dust cloud. Can we say 
more?

We have just a few thoughts to offer concerning this question. It should be possible to 
say something about how $\B$ looks like very close to the junction hypersurface, 
or at least whether it is spacelike there. The answer will be decided by trapped 
surfaces extending partly into the Schwarzschild region. A simple observation we can 
make is that the round trapped surfaces inside Schwarzschild can be ``tilted'' to a 
considerable extent also very close to the event horizon, and still remain trapped. Thus, consider topological spheres at constant $r$ defined by 

\begin{equation} r = r_0 \ , \hspace{8mm} t = cr_0\cos{\theta} \ , \end{equation}

\noindent where $c$ is a constant. Their intrinsic geometry is that defined by the 
intersection of a cylinder with a tilted plane. The point we wish to make is that these 
surfaces are trapped for all values of $c < 1$, independently of the value of $r_0 < 2M$. If the 
junction to the dust cloud is placed at some $t = t_0$ they can (at the expense of some 
effort) be matched smoothly to a surface within the dust cloud. If it can be shown that the 
resulting surfaces can be closed in the Friedmann part they will provide relevant 
information about the location of the boundary, and perhaps suffice to prove that the 
boundary $\B$ is spacelike where it joins the horizon. 
We think it would be interesting to carry such a calculation through. 

%It would nicely 
%complement other investigations, which give conditions on the stress-energy tensor 
%guaranteeing that a black hole spacetime contains a spherically symmetric tube of 
%marginally trapped surfaces which is achronal and asymptotic to the event horizon 
%\cite{Williams}.  
 
We remark that it is easy to show using a perturbation argument as in \cite{BSlang} based on the stability operator for MTS \cite{Lars}, that 
every round sphere on the dashed null cone shown in Fig. \ref{os2}, except the one that lies 
on the event horizon, can be perturbed so that it partly extends into the interior 
of the cone while remaining trapped. However, such arguments do not suffice to 
show that the boundary is spacelike where it meets the horizon. 
 
Another possible line of investigation would be to find past barriers for trapped 
surfaces, lying to the future of the constant Kodama time hypersurface $\Sigma$ 
\cite{BSlang}. In section 5 we showed that the hypersurface $\sigma$ forms such a 
barrier for axially symmetric trapped surfaces reaching the centre. But for more general trapped surfaces 
there is a problem with this. A spherically symmetric spacelike hypersurface 
within the dust cloud is defined by

\begin{equation*} \eta = \eta (\chi) \ . \end{equation*}

\noindent Let the eigenvalues of its second fundamental form be denoted by $(\lambda_1, 
\lambda_2, \lambda_2)$, and its timelike unit normal by $\vec{n}$. A surface of revolution 
within such a hypersurface can be defined by 

\begin{equation*} \theta = \theta (\chi) \ . \end{equation*}

\noindent Let us furthermore assume that this surface is minimal within the hypersurface 
(and that it intersects the matching hypersurface in a circle, where it should be 
continued into Schwarzschild). Then its null expansions are determined by the mean 
curvature vector $\vec{H}$ contracted into the normal vector. For a surface of this kind 
it can be shown that  

\begin{equation} 2\theta_\pm = \vec{H}(n) = \lambda_1 + \lambda_2 - 
\frac{(\lambda_1 - \lambda_2)\theta^{\prime 2}\sin^2{\chi}}{1 - \eta^{\prime 2} + 
\theta^{\prime 2}\sin^2{\chi}} \ . \end{equation} 

\noindent Hypersurfaces whose equatorial cross sections are (locally) marginally 
trapped surfaces -- such as $\sigma$ -- are singled out by the requirement 
$\lambda_1 + \lambda_2 = 0$, and they obey $\lambda_1 > \lambda_2$. But then we 
see that any surface of revolution which is minimal within such a hypersurface will be genuinely trapped whenever $\theta^\prime \neq 0$. Of course we have not shown that such surfaces can be turned into closed trapped surfaces when extended into the Schwarzschild region, but it is already clear that the local argument that shows $\Sigma$ to be a past barrier for all trapped surfaces \cite{BSlang} does not carry over to $\sigma$ in any simple way.   

Finally we want to raise a curious issue. It is known that no observer in a 
pure Schwarzschild black hole can observe a trapped surface in its entirety 
\cite{Iyer}. The same is presumably true for the Vaidya solution, but certainly 
it is not true in the OS model. Can one pinpoint exactly what it takes for a 
trapped surface to be visible? A first suggestion, that outermost stable MTS 
can never be fully observed, fails because an observer falling along the central 
world line in the Oppenheimer-Snyder spacetime can observe some of the stable 
marginally trapped surfaces that form the Schwarzschild part of the event 
horizon, see Fig.\ref{fig:OS1}. Observe, however, that this is not the case for the Vaidya solution as represented in Fig.\ref{fig:trap4}. Even though at the moment we simply do not know the final answer, we conjecture that the difference arises due to the fact that the spherically symmetric marginally trapped tube A3H is timelike in the former case, while it is spacelike in the latter. Other possible Penrose diagrams in spherical symmetry (for instance figures 5 and 6 in \cite{BSlang}) seem to support this idea, even though a proper proof would probably require some effort.

%%%%%%%%%%%%%%%%%%%%%%%%%%%%%%%%%%%%%%%%%%%%%%%%%%%%%%%%%%%%%%%%%%%%%%%%%%
% CONCLUSIONS
%%%%%%%%%%%%%%%%%%%%%%%%%%%%%%%%%%%%%%%%%%%%%%%%%%%%%%%%%%%%%%%%%%%%%%%%%%

\section{Conclusions}
The question that underlies our investigation is: In a dynamical black 
hole, where is the boundary of the region containing trapped surfaces? As in a 
previous investigation of the Vaidya spacetime \cite{BSkort} we focus on a 
special kind of trapped surfaces designed to reach this boundary at the centre 
of spacetime -- although they would probably be overlooked in a numerical simulation 
tied to a specific foliation. There are two major improvements compared to the previous 
investigation:

\begin{itemize}

\item{We have ensured that the surfaces are everywhere differentiable.}

\end{itemize}

\noindent The conditions for this to be true across a matching hypersurface are 
given in the Appendix, and may be of independent interest.

\begin{itemize}

\item{We have made a serious effort to optimize the construction.}

\end{itemize}

\noindent Although a proof escapes us, we believe we have reached the boundary 
of the trapped region at the centre of the Oppenheimer-Snyder dust cloud. Because 
our construction is fully explicit a rigorous temporal upper bound on the location 
of the boundary is achieved. 

Various open issues were discussed in section 6. We find it puzzling, and indeed 
intriguing, that the very simple questions we ask are so difficult to answer.

%ACKNOWLEDGEMENTS

\subsection*{Acknowledgements:}
IB and EJ thank Istvan Racz for encouragement. IB was supported by the Swedish 
Research Council under contract VR 621-2010-4060. JMMS is supported by grants FIS2010-15492 (MICINN), GIU06/37 (UPV/EHU),
P09-FQM-4496 (J. Andaluc\'{\i}a---FEDER) and UFI 11/55 (UPV/EHU).

%%%%%%%%%%%%%%%%%%%%%%%%%%%%%%%%%%%%%%%%%%%%%%%%%%%%%%%%%%%%%%%%%%%%%%%%%
% APPENDIX
%%%%%%%%%%%%%%%%%%%%%%%%%%%%%%%%%%%%%%%%%%%%%%%%%%%%%%%%%%%%%%%%%%%%%%%%%

\appendix
\section{Appendix}
In this Appendix we consider the question of how to deal with surfaces in spacetimes $(M,g)$ that are the result of a matching between two, previously given, known spacetimes $(M^+,g^+)$ and $(M^-,g^-)$ across a (timelike for definiteness) hypersurface $\E$. 

As is often the case -- and sometimes unavoidable -- the explicit coordinates used to describe such matched spacetimes $(M,g)$ consist of two different, unrelated sets: one corresponding to the $+$-manifold $M^+$, the other corresponding to $M^-$. Even though one cannot even ask the question if the metric components, say, are differentiable functions across $\E$ in such coordinates, this leads to no conceptual difficulties because there are theorems \cite{I,CD,MS} ensuring that, provided the proper matching conditions hold, there exists another coordinate system on any neighbourhood $U(p)\subset M$ of $p\in\E$ such that the metric components are actually $C^1$ functions in this new set. These coordinates are called admissible \cite{L,I}, and are usually not constructed explicitly. Actually, in most cases their expressions will be rather difficult, if not impossible, to get in terms of explicit functions. 

All this is well understood. However, we want to consider surfaces (co-di\-men\-sion two submanifolds) that cross the matching hypersurface $\E$. How can one be sure that a given surface is actually differentiable, without undesired ``corners'' or ``spikes'', and such that extrinsic quantities -- as for instance the expansions -- do not have jumps? We are going to provide a simple method to deal with this question without any knowledge of the admissible coordinates. To this end, a very brief summary of the junction conditions is in order, see \cite{Exact} section 3.8 for a summary.

Let $(M^{\pm},g^{\pm})$ be two smooth spacetimes with respective metrics $g^{\pm}$. Assume that there are corresponding timelike hypersurfaces $\E^{\pm}\subset M^{\pm}$ which bound the regions $V^{\pm}\subset M^{\pm}$ on each $\pm$-side to be matched. These two hypersurfaces are to be identified in the final glued spacetime, so they must be diffeomorphic. The glued manifold is defined as the disjoint union of $V^{+}$ and $V^{-}$ with diffeomorphically related
points of $\E^{+}$ and $\E^{-}$ identified. The matching depends crucially on the particular diffeomorphism used to identify $\E^{+}$ with  $\E^{-}$, and we assume that this has already been chosen and is given, so that the matching hypersurface is uniquely and well defined. Henceforth, this identified
hypersurface will be denoted simply by $\E$. A necessary requirement to build a spacetime with at least a continuous metric is that the first fundamental forms $h^{\pm}$ of $\E$ calculated on both sides agree, because then
there exists a unique $C^{1}$ atlas on $M$, which induces the $C^{\infty}$ structures on $M^{\pm}$ and such that there is a metric extension $g$ defined on the entire manifold that coincides with
$g^{\pm}$ in the respective $V^{\pm}$ and is continuous \cite{CD,MS}.

In practice, one is given two spacetimes $(M^{\pm},g^{\pm})$ and thus two sets of local coordinates $\{x^\mu_{\pm}\}$ {\em with no relation whatsoever}. Hence, one has two parametric expressions $x_{\pm}^\mu =x_{\pm}^\mu (\xi^a)$ of $\E$, one for each imbedding into each of $M^\pm$. Here $\{\xi^a\}$ are intrinsic local coordinates
for $\E$, Greek indices run from 0 to 3, while small Latin indices run from 1 to 3. For each $\pm$ sign, the three vector fields $e^{\mu}_{b}=\partial x_{\pm}^\mu/\partial\xi^{b}$ are assumed to be linearly independent at any $p\in \E$ and are tangent to $\E$. The agreement of the two $(\pm)$-first fundamental forms amounts to the equalities on $\E$
$$
h^+_{ab} = h_{ab}^- ,\hspace{1cm}
h^\pm_{ab} \equiv g_{\mu\nu}^\pm (x(\xi))\frac{\partial x_\pm^\mu}{\partial \xi^a}\frac{\partial x_\pm^\nu}{\partial \xi^b} \, .
$$
In other words, the tangent vector fields have equal scalar products on $\E$ from both sides. 
Denote by $n^\pm_\mu$ two unit normals to $\E$, one for each side. They are fixed up to a sign by the conditions
$$
n^\pm_\mu \frac{\partial x_\pm^\mu}{\partial \xi^a}=0, \hspace{1cm} n^\pm_\mu n^{\pm\mu}=1
$$
and one must choose $n^-_\mu$ pointing outwards from $V^-$ and $n^+_\mu$ pointing towards $V^+$ -- or the other way round. The two bases on the tangent spaces 
$$\{n^{+\mu},\frac{\partial x_+^\mu}{\partial \xi^a}\}\quad \mbox{and} \quad \{n^{-\mu},\frac{\partial x_-^\mu}{\partial \xi^a}\}
$$
are then identified, so that one can drop the $\pm$. Observe, however, that this identification is usually only abstract, as in practice one still uses both bases, each on its own coordinate system, to do actual calculations.

The complete set of matching conditions is then obtained by requiring that the Riemann tensor components have no Dirac-delta-type terms \cite{CD,MS}, and this amounts to demanding that the second fundamental forms of $\E$, as computed from both $\pm$-sides, agree on $\E$ \cite{I,D}, that is to say
$$
K^{+}_{ab}=K^{-}_{ab},  \hspace{1cm} K^{\pm}_{ab}=-n^{\pm}_{\mu}\left(\frac{\partial^{2} x^{\mu}_{\pm}}{\partial\xi^a\partial\xi^{b}}+\Gamma^{\pm\mu}_{\alpha\beta}(x(\xi)) \frac{\partial x_\pm^\alpha}{\partial \xi^a}\frac{\partial x_\pm^\beta}{\partial \xi^b}\right) \, .
$$

We are now in position to answer the question we asked. Assume, thus, that the above procedure has been performed and we have a properly matched spacetime $(M,g)$ which, nevertheless, is presented to us with the two portions $V^{\pm}$ and the two metrics $g^{\pm}$ described in their original (and unrelated) $\pm$-coordinate systems. Imagine that we wish to describe an imbedded surface $S$ of sufficient differentiability in this space-time -- so that, in particular, the null expansions on $S$ are continuous. How to proceed? 

Consider for a moment that we have built an admissible coordinate system $\{x^{\mu}\}$ around a point $p\in \E$. This means that the metric $g$ is $C^{1}$ across $\E$ in these coordinates, and that $g$ coincides with -- i.e., is isometric to -- $g^{+}$ on $V^{+}$, and $g^{-}$ on $V^{-}$. The surface $S$ would then be described in parametric form by
$$
x^{\mu}=\Phi^{\mu}(\lambda^{B})
$$
where the functions $\Phi^{\mu}$ are sufficiently differentiable (say, at least $C^{3}$) and $\lambda^{B}$ are local intrinsic coordinates in $S$. Now, capital Latin letters $A,B,\dots =2,3$. This implies that the vector fields tangent to $S$
$$
\frac{\partial \Phi^{\mu}}{\partial \lambda^{B}}
$$
possess components that are $C^{2}$ functions of the $\lambda^{B}$, and the components of the first fundamental form of $S$
$$
\gamma_{AB}= g_{\mu\nu}(\Phi(\lambda)) \frac{\partial \Phi^{\mu}}{\partial \lambda^{A}}
\frac{\partial \Phi^{\nu}}{\partial \lambda^{B}}
$$
are therefore $C^{1}$ functions of the $\lambda^{B}$. With respect to the normal one-forms, any normal $N_{\mu}$ to $S$ is defined by the condition
$$
N_{\mu}\frac{\partial \Phi^{\mu}}{\partial \lambda^{B}}=0
$$
and therefore its components $N_{\mu}(\lambda^{B})$ can be chosen to be $C^{2}$ as functions of the $\lambda^{B}$. In particular, this will be the case for the two independent null normal one-forms. Observe, however, that the contravariant components $N^{\mu}$ will, in general, be just $C^{1}$ functions. Despite this, the components of the second fundamental form relative to any normal $N_{\mu}$,
$$
K_{AB}(N)=-N_{\mu}\left(\frac{\partial^{2} \Phi^{\mu}}{\partial\lambda^A\partial\lambda^{B}}+\Gamma^{\mu}_{\alpha\beta}(\Phi(\lambda)) \frac{\partial \Phi^\alpha}{\partial \lambda^A}\frac{\partial \Phi^\beta}{\partial \lambda^B}\right),
$$
are continuous functions of the $\lambda^{B}$, because so are the Christoffel symbols as functions of the admissible $x^{\mu}$. Finally, the normal connection one-form $s_{A}$ on $S$, also called the third fundamental form of $S$, has the following components 
$$
s_{A}=-m_{\mu}\left(\frac{\partial u^\mu}{\partial \lambda^{A}}+\Gamma^{\mu}_{\alpha\beta}(\Phi(\lambda))\frac{\partial\Phi^{\alpha}}{\partial\lambda^{A}}u^{\beta}\right)
$$
in terms of any orthonormal couple $u_{\mu},m_{\mu}$ of normal one-forms
$$
-u_{\mu}u^{\mu}=m_{\mu}m^{\mu}=1, \quad u_{\mu}m^{\mu}=0.
$$
It follows that the two summands between brackets of the expression for $s_{A}$ are continuous functions of the $\lambda^{B}$ and thus, so are $s_{A}$.

The above has been deduced using an admissible coordinate system that we will not generally have at hand. Still the conclusions reached, namely the differentiability of the first fundamental form and the continuity of the second and third fundamental forms as functions of the $\lambda^{B}$, are invariant and can be enforced without the use of the admissible coordinates. This follows because all these geometrical quantities have components that are expressed in terms of the intrinsic local coordinates $\lambda^{B}$ on $S$ so that, as long as we can provide expressions for these on both sides as such functions, their continuity or differentiability can be explicitly required. 

This works as follows. In the practical situation we will have to describe the surface $S$ as composed by a piece $S^{+}$ imbedded in $V^{+}$, another piece $S^{-}$ imbedded in $V^{-}$, and such that both pieces intersect the matching hypersurface $\E$ at the same set 
\be
S^{+}\cap \E =S^{-}\cap \E \equiv S|_{\E}. \label{SinE}
\ee
The imbeddings will each be described by corresponding parametric functions 
$$
x^{\mu}_{\pm}=\Phi^\mu_{\pm}(\lambda^{B})
$$
so that, first of all, we need to find the explicit expressions that solve the indispensable condition (\ref{SinE}), that is to say, we need to find the solution to the two systems of equations
$$
x^\mu_\pm (\xi^b) = \Phi^\mu_{\pm}(\lambda^{B})\, .
$$
This solution exists, and actually agree on both sides, if the surface $S$ does meet the matching hypersurface $\E$ and $S|_\E$ is well defined. Let such a solution be described by the explicit functions and constraints
\be
\xi^b =\xi^b (\lambda^B_0),Ê\quad \quad {\cal F}_{\Omega}(\lambda^{B})=0. \label{xioflambda}
\ee
Here, $\lambda^{B}_0$ denote the values of $\lambda^{B}$ at the intersection $S|_{\E}$, that is, the solutions to the constraint equations ${\cal F}_{\Omega}=0$. These may depend on the matching hypersurface $\E$ and on the spacetimes $M^{\pm}$.

We can compute the first fundamental form of $S$ on each of its two pieces $S^{\pm}$ as
$$
\gamma^{\pm}_{AB}= g^{\pm}_{\mu\nu}\left(\Phi_{\pm}(\lambda)\right) \frac{\partial \Phi_{\pm}^{\mu}}{\partial \lambda^{A}}
\frac{\partial \Phi_{\pm}^{\nu}}{\partial \lambda^{B}}\, .
$$
Then, the condition that $\gamma_{AB}$ are differentiable implies that we must require on $S_{\E}$
\be
\left.\gamma^{+}_{AB}\right|_{S_{\E}}= \left.\gamma^{-}_{AB}\right|_{S_{\E}}, \hspace{1cm} \left.\frac{\partial\gamma^{+}_{AB}}{\partial\lambda^{C}}\right|_{S_{\E}}=\left.\frac{\partial\gamma^{-}_{AB}}{\partial\lambda^{C}}\right|_{S_{\E}} \, .\label{1st}
\ee
Observe that taking the values of these components at $S_{\E}$ amounts to setting $\lambda^B =\lambda^{B}_0$.

In order to deal with the rest of the continuity conditions, we must first of all identify properly, on the set $S|_{\E}$, the normal one-forms that are defined on both pieces $S^{\pm}$ of $S$. This can be done by computing, on each side, their scalar products with the identified bases on $\E$.  
We must thus require for any normal one-form $N_\mu$ to $S$
\be \label{normals}
\begin{aligned}
& N^{+}_{\mu}(\lambda_0)\,  n_{+}^{\mu}(\xi(\lambda_0))=N^{-}_{\mu}(\lambda_0)\, n_{-}^{\mu}(\xi(\lambda_0)), \\
& N^{+}_{\mu}(\lambda_0) \frac{\partial x^{\mu}_{+}}{\partial\xi^{a}}(\xi(\lambda_0))= N^{-}_{\mu}(\lambda_0) \frac{\partial x^{\mu}_{-}}{\partial\xi^{a}}(\xi(\lambda_0)).
\end{aligned}
\ee
Note that, given the normal one-form on one side, these can be seen as equations determining the normal one-form on the other side if the surface $S$ is well defined. Once we have the normals properly identified, we can compute the second and third fundamental forms on both sides by using the corresponding $\pm$-objects, in other words
\bean
K^\pm_{AB}(N)=-N^\pm_{\mu}\left(\frac{\partial^{2} \Phi_\pm^{\mu}}{\partial\lambda^A\partial\lambda^{B}}+\Gamma^{\pm\mu}_{\alpha\beta}(\Phi_\pm(\lambda)) \frac{\partial \Phi_\pm^\alpha}{\partial \lambda^A}\frac{\partial \Phi_\pm^\beta}{\partial \lambda^B}\right),\\
s^\pm_{A}=-m^\pm_{\mu}\left(\frac{\partial u_\pm^\mu}{\partial \lambda^{A}}+\Gamma^{\pm\mu}_{\alpha\beta}(\Phi_\pm(\lambda))\frac{\partial\Phi_\pm^{\alpha}}{\partial\lambda^{A}}u_\pm^{\beta}\right),
\eean
and then we must require, by letting $\lambda^{B}=\lambda^{B}_0$,
\be
\left.K^+_{AB}(N)\right|_{S_{\E}}=\left.K^-_{AB}(N)\right|_{S_{\E}} , \quad\quad  \left.s^+_A\right|_{S_{\E}} = \left.s^-_A\right|_{S_{\E}} \, .\label{2ndand3rd}
\ee

In summary, given the discussion above in admissible coordinates, the set of conditions that we must require on $S$ are
\begin{enumerate}
\item the existence of a solution (\ref{xioflambda}) that defines the set (\ref{SinE}) unambiguously,
\item the proper identification of normal one-forms on $S|_{\E}$ according to (\ref{normals}),
\item conditions (\ref{1st}) to allow for the differentiability of the first fundamental form on $S|_{\E}$,
\item and finally conditions (\ref{2ndand3rd}) to comply with the continuity of the second and third fundamental forms.
\end{enumerate}
Clearly, these conditions are necessary to have a well-defined surface $S$ without corners, with no jumps in the expansions, etc. They are actually sufficient too, because if the surface $S$ had a corner, or if its null expansions had a jump, etc., then some of them would not hold.

As a final comment, we want to remark that the construction carried out in a previous paper \cite{BSkort} was not completely correct in this sense, because the advanced null coordinates used there are not admissible (the metric has jumps in some of their first derivatives) while the built trapped surface was given parametrically in terms of those coordinates. This can actually be noticed because the expansions had a jump across the matching hypersurface\footnote{Even though the matching hypersurface in \cite{BSkort} is null, a similar treatment as the one herein presented can be carried out along the lines of \cite{CD,MS}.}. Nevertheless, the surface could easily be smoothed out and the construction would still work.

%%%%%%%%%%%%%%%%%%%%%%%%%%%%%%%%%%%%%%%%%%%%%%%%%%%%%%%%%%%%%%%%%%%%%%%%%%
% REFERENCES
%%%%%%%%%%%%%%%%%%%%%%%%%%%%%%%%%%%%%%%%%%%%%%%%%%%%%%%%%%%%%%%%%%%%%%%%%%

\end{document}